\documentclass[aps,prb,showpacs,twocolumn]{revtex4-1}

\usepackage{ulem}
\usepackage{graphicx}
\usepackage{CJK}
\usepackage{amssymb}
\usepackage{epsfig}
\usepackage{graphicx}
\usepackage{subfigure}
\usepackage{mathrsfs}
\usepackage{verbatim}
\usepackage{rotating}
\usepackage{longtable}
\usepackage{amsmath}
\usepackage{array,color}
\usepackage{dcolumn}
\usepackage{bm}

\begin{document}
\title{Driver at 10 MJ and 1 shot$/$30min for inertial confinement fusion at high gain:
efficient, compact, low-cost, low laser-plasma instabilities, beam-color selectable from 2$\omega$$/$3$\omega$$/$4$\omega$$/$, applicable to multiple laser fusion schemes}
\begin{CJK*}{GB}{gbsn}

\author{Zhan Sui (ËåÕ¹)}\email{lqling@vip.163.com}
\affiliation{Shanghai Institute of Laser Plasma, China Academy of Engineering Physics, Shanghai 201800, China}

\author{Ke Lan (À¶¿É)}
\affiliation{Institute of Applied Physics and Computational Mathematics, Beijing 100094, China}
\date{\today }

\begin{abstract}
The ignition at the National Ignition Facility (NIF) set off a global wave of research on the inertial fusion energy (IFE). However, IFE requires a necessary target gain G of 30-100, while it is hard to achieve the fusions at such high gain with the energy, configuration, and technical route of the NIF. We will present a conceptual design for the next generation laser driver of 10 MJ, 2 $\sim$ 3 PW at 3$\omega$ (or 2$\omega$, then the energy and power can be higher), and 1 shot$/$30min, which is efficient, compact, low-cost, low laser-plasma instabilities, applicable to multiple laser fusion schemes, and aiming for G $>$ 30.
\end{abstract}

\maketitle

\section{Introduction}
\begin{figure}[htbp]
\includegraphics[width=0.4\textwidth]{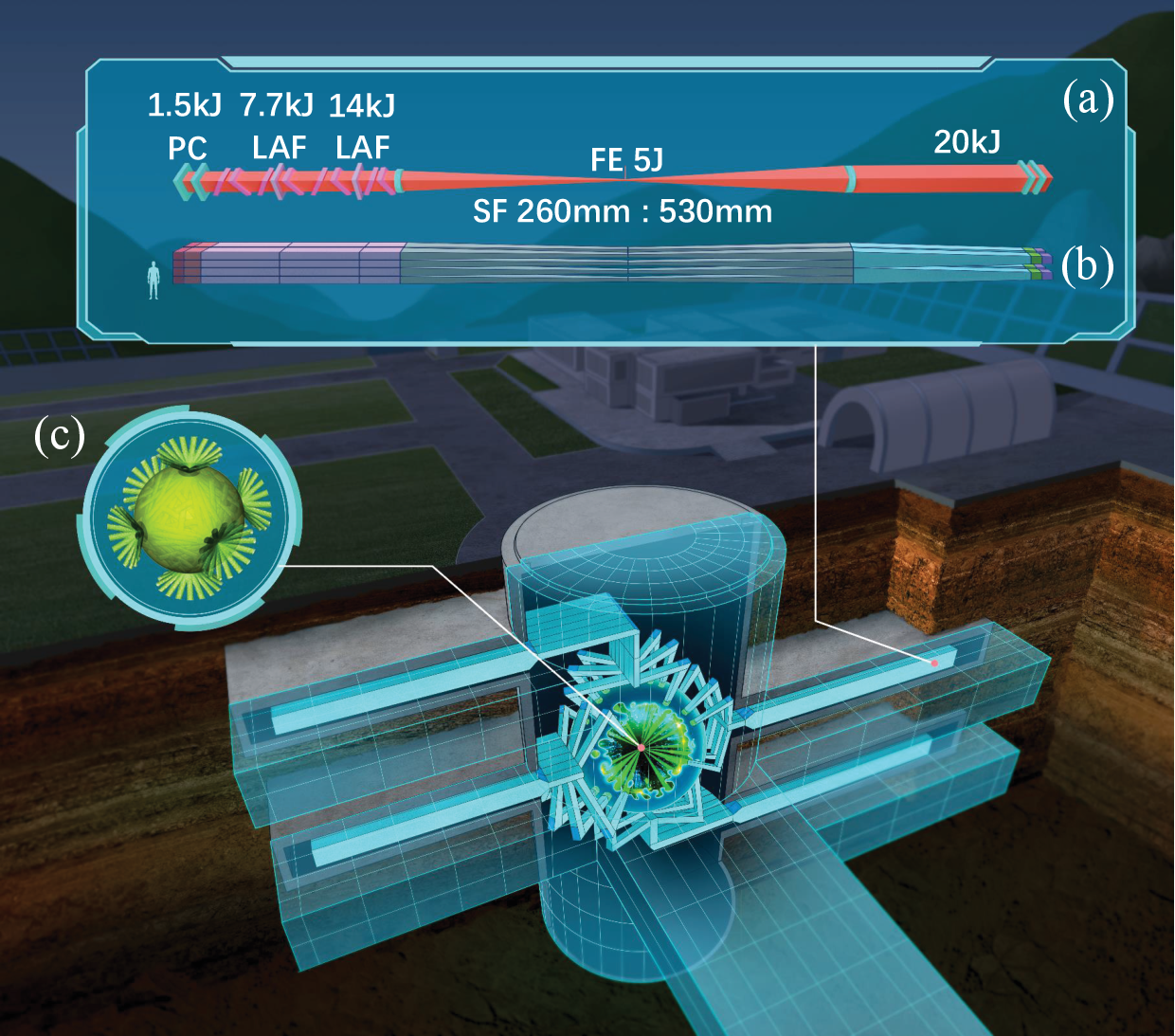}
\caption{(Color online)
Schematic of the general layout of the laser driver with its target chamber. Insets: (a) the arrangement of a laser beam, which is injected at FE with an initial energy of 5 J, reflected at PC with an amplified energy of 1.5 kJ,  then  amplified in the main amplifier and filtered by LAFs, and finally delivered at SF with an energy of 20 kJ and a beam size expansion from 260 mm to 530 mm; (b) a beamlet of 4 beams at 3$\omega$; (c) octahedral spherical hohlraum and its ideal laser arrangement with 16 beamlets per end.
Here, FE is the frond end, LAF is the laser filtering, PC is the Pockels cell, and  SF is the Spatial Filter. The figure on the left of inset (b) shows the scale of the laser driver.}
\label{Fig:Fig}
\end{figure}

Laser driven inertial fusion is a highly promising approach toward fusion energy, which has been a quest of human beings for more than a half century. The National Ignition Facility (NIF) at Lawrence Livermore National Laboratory  achieved target gain G $>$ 1 in the end of 2022 \cite{Abu-Shawareb2024PRL} and set a new fusion yield record of 5.2 MJ in February, 2024 \cite{5dot2MJyield}, which successfully demonstrated the feasibility of laboratory laser fusion.  However, it still far below the target gain G of 30 to 100 required by the inertial fusion energy (IFE) \cite{MTV}. In fact, limited by its configuration \cite{Lan2022MRE}, energy \cite{NNSA2015}, and the technical route identified in the 1980s \cite{Haynam2007, Manes2016}, it is hard for the NIF to reach such high gain. Hence, a driver with novel laser technologies is mandatory in the path toward IFE. In this paper, we will give a conceptual design for an efficient and low-cost laser driver of 1 shot/30min, aiming for G $>$ 30. A schematic  is shown in Fig. \ref{Fig:Fig}, and the main technical specifications are shown in Table 1. In the following, we will address the novel technologies  to improve the energy storage efficiency, extraction efficiency, flux, volume, etc. for this driver.

\begin{table}[htbp]
\caption{Main technical specifications.}
\begin{tabular}{|c|c|}
\hline
Laser energy & $>$ 10 MJ \\
\hline
Laser power &  2 $\sim$ 3 PW \\
\hline
Laser wavelength & beam-color selectable
\\
  &  from 2$\omega$$/$3$\omega$$/$4$\omega$   \\
\hline
Waist diameter  & 1000 $\mu$m \\
\hline
Main pulse width & 3 $\sim$ 5 ns \\
\hline
Repetition rate &  1 shot/30min \\
\hline
F Number & 10 $\sim$ 15 \\
\hline
End number & 6  \\
\hline
Beamlet number per end & $N_Q$ = 16\\
\hline
Beam number per beamlet & 4  \\
\hline
Driving mode & octahedral configuration \\
\hline
Shooting angle  &  all  beamlets have the same \\
  for details:  & incidence angle $\theta_L$     \\
  Refs. [\citenum{Lan2014POP, Lan2022MRE}] &  ranging from 50$^\circ$ to 60$^\circ$,   \\
  &  and the same initial  azimuthal \\
&  angle $\phi _{L0}$ ranging  \\
&  from $0$ to $2\pi/{4N_Q}$. \\
\hline
Cost &  $\le$ 20 billion RMB \\
\hline
\end{tabular}
\end{table}

\section{Key technologies}

\noindent {\bf 1. Multi-front-end technology}

Laser-plasma instabilities (LPIs) is one of the main obstacles of the NIF to achieve high gain. In our novel design, each laser beam has an independent front-end, i.e. one distributed feedback (DFB) oscillator per beam line. Then, the regulation freedom can be improved because each laser beam has no fixed phase relationship, and therefore, the incoherent superposition can be realized at the target point. This is conducive to the beam smoothing and can effectively suppress LPIs. Moreover, a super light spring of incoherence in all dimensions of time, space, and angle can be used to further suppress LPIs \cite{incoherent2023MRE}. Thus, it is possible for the driver to work also at a frequency lower than 3$\omega$, for a higher energy conversion efficiency and a higher damage threshold for the optic components in the final optic assembly.

\noindent {\bf 2. Near-field spatial separation amplification of pre- and main pulse}

In the ignition target designs, the laser pulse consists of pre-pulse and main pulse, and the power of pre-pulse is usually much lower than that of the main pulse. This makes the conversion efficiency of the pre-pulse part very low during the frequency conversion process, leading to an overall frequency conversion efficiency of the ignition pulse about 30$\%$ lower than that of the square-wave pulse \cite{Manes2016}. In our novel design, the pre- pulse and main pulse are spatially separated in the amplifier by adopting the spatiotemporal shaping. Furthermore, the pre-pulse  corresponds to a smaller aperture, while the main pulse corresponds to the rest. This can be used to adjust the power densities of the pre-pulse and the main pulse to be equivalent, so as to improve the overall conversion efficiency. The superposition effect of pre-pulses and main pulses at the target point can be the same as that of the traditional amplification method.

\noindent {\bf 3. High fluence amplification employed laser material with low emission cross section, long fluorescence lifetime, and high energy storage}

Most of existing laser drivers use neodymium glass as the gain medium, which has a large radiative transition probability and a high gain due to its short fluorescence lifetime and large emission cross section, however, it has a relatively small energy storage. In our novel design, the driver uses the new sensitizing doped laser material, which has a fluorescence lifetime ( $\sim$650 $\mu$s), low emission cross section ($\sim$ 2 $\times$ 10 $^{-20}$ cm$^2$) and high stored energy (0.5 J/cm$^3$) as the main amplifier gain material. In addition, a high-flux and multi-pass amplification is used to effectively extract the stored energy. This type of amplifier is expected to achieve a high flux output of 30 J/cm$^2$.

\noindent {\bf 4. Water-cooled xenon lamp with annular section and fluorescent conversion separator material}

As known, the operational efficiency of high-power laser devices is relatively low, only about 3 - 4 shots per day. In the novel design, we use a xenon lamp with an annular section as the pump, which  fluorescence can be converted into an absorption band and absorbed by the laser medium by using a fluorescence conversion baffle material. Hence, the pumping light-laser conversion efficiency can be remarkably improved. Meanwhile, the xenon lamp is subjected to water cooling, so that the heat can be taken away in time and the laser medium can be prevented from heating up continuously. As a result, the emission interval can be effectively shortened, and a shooting capability of 30 minutes per emission can be expected. This technology can improve the radiation efficiency of the large diameter xenon lamp and the absorption efficiency of gain medium, leading to a much higher energy storage efficiency of disk amplifier and a much shorter emission interval.

\noindent {\bf 5. Near-field multi-pass split-plate amplifier based on angle-sensitive film}

The energy extraction efficiency is crucial important for a driver. In order to improve the extraction efficiency, we design a near-field multi-pass amplifier. The amplifiers have a certain splitting angle, and the film layers of surfaces are designed to be angle-sensitive. By this way, the pulse has a high reflectivity on the surfaces in the first two paths of transmission, and the angle corresponding to the third path has a high transmissivity \cite{Spaeth2016}. Thus, the actual transmission distance of the pulse passing through the amplification medium is three times the thickness of the medium. In other words, it is a three-pass amplification, in which the pulse has nearly equal fluence in one amplifier medium while the integral fluence is three times enhanced. This amplification mode has a simple system structure but a high extraction efficiency.

\noindent {\bf 6. Two-pass amplification configuration combined with near-field three-pass amplification}

Combined with a near-field three-pass amplification, the main amplifier is equipped with a two-pass amplification configuration, and the amplification of 48 equivalent pieces can be realized by adopting 8 pieces of gain media. This system has a compact structure (with a length about 20 m), a large integral flux, a balanced flux of each piece of gain medium, and a high extraction efficiency, which can meet the image transmission of the system when combined with the traditional spatial filter.

\noindent {\bf 7. Spatial filtering technology based on angular spectrum sensitive nonlinear crystal}

In current laser technology, it takes the low-gain high-fluency amplification. In this case, the B integral of the system is severe, and a spatial filtering is required to suppress the increase of B integral \cite{Haynam2007}. However, the traditional spatial filters generally have a long size and require a vacuum assembly, which cost is expensive. In our design, we use the nonlinear near-field filtering method in the designed single or cascaded nonlinear crystals, then the spatial high-frequency components of pulses can be converted to the second-harmonic wave and filtered in the following amplification processes. In addition, the low-frequency components of pulses have a low conversion efficiency, so the spatial filtering can be realized. Especially, the nonlinear spatial filtering can be conveniently inserted into the amplifier, resulting in a more compact structure of the laser driver with significant space and cost savings.

\noindent {\bf 8. Beam combination system based on the non-collinear nonlinear frequency conversion}

As known, the beam number of laser driver is restricted because the solid angle of the target chamber is fixed. In our  design, the two fundamental- frequency beams of light are combined into one beam at third-harmonic frequency, through non-collinear second-harmonic and sum-frequency conversion. Furthermore, the beam number can be doubled by using the non-collinear beam combination scheme, and the output pulse can be ensured to have sufficient brightness with a small F number under the condition that the load capacity of the device is constant.

\noindent {\bf 9. New Measurement and Control Technology}

Compared with the current advanced control technologies, the existing control system used in the laser drivers lags far behind. In our design, it will employ the centralized control system based on global wireless Internet of Things (IOT), together with a high-resolution and high-stability time synchronization system based on high- speed communication technology, and a full-system control based on artificial intelligence (AI) and miniaturized measurement package.

\noindent {\bf 10. General layout: semi-underground centered on the target chamber}

According to our above technical schemes, the space size of the driver can be greatly reduced, and also the underground layout becomes possible. The general layout is semi- underground centered on the target chamber, which has six laser shooting ends \cite{Lan2014POP}, with 16 beamlets per  end and 4 beams per beamlet, guided into the target chamber, as shown in Fig. \ref{Fig:Fig}. It can reduce the cost of construction and environmental control.

\section{Summary}
We have presented a conceptual design for the next generation laser driver of 10 MJ, 2 $\sim$ 3 PW  and 1 shot/30min, with the ideal laser arrangement of the octahedral spherical hohlraum approach \cite{Lan2014POP, Craxton2020DPP, Marangola2021}, which is applicable to multiple schemes for IFE route choice and feasibility demonstration.
The highly symmetric radiation drive at all times and for all spectra naturally generated inside an octahedral spherical hohlraum
is mandatory in achieving a fusion of high gain with the delicate capsule designs \cite{Haan2011POP, Ren2018POP, Qiao2021PRL}.
Under a 10 MJ drive, it can be expected to achieve G $\sim$ 30 by assuming a burn depletion \cite{MTV} $\Phi$ $\sim$ 45$\%$ of a 2 mg DT fuel by indirect-drive, and G  $\sim$ 100 by assuming $\Phi$ $\sim$ 30$\%$ of a 10 mg DT fuel by direct-drive.
Here, we consider a lower $\Phi$ for direct-drive due to its worse asymmetry aroused by the laser imprinting and beam crossing.
Detail designs of such  targets are to be presented in our forthcoming publications.
As in Table I, the beam-colors of the novel driver are selectable from  2$\omega$$/$3$\omega$$/$4$\omega$$/$ among beamlets,  which means that each laser beam can simultaneously and independently work at either 2$\omega$, 3$\omega$, or 4$\omega$ for various physics purposes \cite{Lan2017POP, CYH2018POP}. This can be realized  via nonlinear frequency conversion process  by removing or changing the crystals.
Compared with the prior art, our novel laser driver design can remarkably improve the energy storage efficiency, extraction efficiency, flux, repetition, and meanwhile, greatly reduce its volume, metal components and cost.

{\bf ACKNOWLEDGMENTS}
The authors appreciate Dr. Xiaohui Zhao of Shanghai Institute of Laser Plasma, Dr. Yongsheng Li, Dr. Hui Cao,  Dr. Yaohua Chen, and Dr. Xiumei Qiao of Institute of Applied Physics and Computational Mathematics in Beijing for their beneficial helps.
This work is supported by the National Natural Science Foundation of China (Grant No. 12035002).
\end{CJK*}

\end{document}